\documentclass[12pt]{article}
\usepackage{epsf,rotate,amssymb}
\topmargin by -\headsep \textheight 8.9in \oddsidemargin 0pt
\evensidemargin \oddsidemargin \marginparwidth 0.5in \textwidth
6.5in 

\newcommand{\bi}{\begin{itemize}}
\newcommand{\ei}{\end{itemize}}

\newcommand{\dlmax}{d_{l\mbox{max}}}

\newcommand{\Vcalhat}{\hat{\cal V}}
\newcommand{\Xcal}{{\cal X}}

\newcommand{\Acal}{{\cal A}}
\newcommand{\Vcal}{{\cal V}}

\newcommand{\Bcal}{{\cal B}}
\newcommand{\Zcal}{{\cal Z}}

\newcommand{\vhat}{\mbox{$\hat{v}$}}

\newcommand{\Ihat}{\hat{I}}
\newcommand{\Itil}{\tilde{I}}

\newcommand{\conv}{\mbox{conv}}

\newcommand{\Scal}{{\cal S}}

\newcommand{\E}{\mbox{E}}

\newcommand{\be}{\begin{equation}}
\newcommand{\ee}{\end{equation}}
\newcommand{\bea}{\begin{eqnarray}}
\newcommand{\eea}{\end{eqnarray}}
\newcommand{\beann}{\begin{eqnarray*}}
\newcommand{\eeann}{\end{eqnarray*}}

\newtheorem{theorem}{Theorem}[section]
\newtheorem{corollary}[theorem]{Corollary}
\newtheorem{lemma}[theorem]{Lemma}

\newtheorem{remark}[theorem]{Remark}

\newtheorem{definition}[theorem]{Definition}

\renewcommand{\theequation}{\arabic{section}.\arabic{equation}}
\renewcommand{\thetable}{\arabic{section}.\arabic{table}}
\renewcommand{\thefigure}{\arabic{section}.\arabic{figure}}

\newcommand{\Section}[1]{\section{#1}
\setcounter{equation}{0}
\setcounter{figure}{0}
\setcounter{table}{0}}

\title{Alphabet Sizes of Auxiliary Variables\\ in Canonical Inner Bounds}

\author{Soumya Jana\\
        Department of Electrical and Computer engineering\\
        University of Illinois at Urbana-Champaign\\
        Email: {\tt jana@uiuc.edu}}
\begin{document}

\maketitle

\thispagestyle{plain}
\pagestyle{plain}

\baselineskip=1.25\normalbaselineskip
\renewcommand{\baselinestretch}{1.4}

\renewcommand{\baselinestretch}{1.5}

\begin{abstract}
Alphabet size of auxiliary random variables in our canonical description is derived. Our analysis improves upon estimates known in special cases, and generalizes to an arbitrary multiterminal setup. The salient steps include decomposition of constituent rate polytopes into orthants, translation of a hyperplane till it becomes tangent to the achievable region at an extreme point, and derivation of minimum auxiliary alphabet sizes based on Caratheodory's theorem.
\end{abstract}

\Section{Introduction}
\label{sec:intro}

The central question in Shannon theory of source coding is the characterization of achievable regions in information-theoretic terms. Historically, simple information-theoretic (so-called `single-letter') descriptions were shown to completely characterize the achievable regions of certain problems, such as Shannon's lossless and lossy coding problems \cite{ShanLL,Shannon}, the Slepian-Wolf problem \cite{SW}, the Wyner-Ahlswede-K\"{o}rner problem \cite{Wyner,AhlKor}, the Wyner-Ziv problem \cite{WZ}, and the Berger-Yeung problem \cite{BY}. Specifically, coincident inner and outer bounds have been found for the aforementioned problems. However, in certain other source coding problems, including the Berger-Tung and the partial side information problems \cite{BT,Upper}, coincident inner and outer bounds have not been found. In this paper, we shall consider a general class of inner bounds, which we call {\em canonical}, and which may or may not be tight \cite{isit08}. For example, our bound coincides with known descriptions in aforementioned solved problems, as well as with Berger-Tung bound known for the Berger-Tung and the partial side information problems. Further, unlike earlier attempts at unification, such as by Csisz\'{a}r and K\"{o}rner \cite{CsisKor}, and Han and Kobayashi \cite{HanKob}, our canonical bound brings both lossless and lossy coding under the same framework. Moreover, our bound is tight for (hence solves) a large class of multiterminal problems \cite{itw07}, generalizing the longstanding single-helper problem \cite{CsisKor}. However, at present we shall not focus on conditions for tightness. Instead we shall analyze an aspect that has historically received very little attention. Note that our inner bounds involve certain auxiliary variables $\{Z_k\}$ with alphabets $\{\Zcal_k\}$ (the notation is made precise subsequently). Alphabet sizes $\{|\Zcal_k|\}$ play an important role in practical computation, and hence understanding of the inner bounds (see, e.g., \cite{GuEffros,GuJanaEffros}). The available results generally estimate $|\Zcal_k|\le|\Xcal_k|+$constant, where $\Xcal_k$ is the given alphabet of the source $X_k$ associated with the auxiliary variable $Z_k$, and the constant is one or greater. In this paper, we shall derive a tight bound $|\Zcal_k|\le|\Xcal_k|$ of such alphabets, thereby, facilitating computation.

As alluded earlier, in different contexts $|\Zcal_k|$ has been estimated within a constant factor of $|\Xcal_k|$. For example, we know $|\Zcal_k|\le |\Xcal_k|+2$ for the Wyner-Ahlswede-K\"{o}rner problem \cite{Wyner,AhlKor}, $|\Zcal_k|\le |\Xcal_k|+1$ the Wyner-Ziv problem \cite{WZ}, and $|\Zcal_k|\le |\Xcal_k|+2$ for the Berger-Yeung problem \cite{BY}. In those problems, there is only one auxiliary variable, and a rate-distortion orthant is varied to create the desired inner bound (which equals the achievable region). In contrast, the Berger-Tung region involves two auxiliary variables, and is created by varying a convex core region, which is more complicated than an orthant \cite{BT}. So far, there exists no rigorous analysis of the alphabet size in this case, but estimates vary between $|\Zcal_k|\le |\Xcal_k|+1$ and $|\Zcal_k|\le |\Xcal_k|+2$. In an earlier paper \cite{itw07}, we gave an estimate of $|\Zcal_k|\le |\Xcal_k|+M$ for the general $M$-terminal single-helper problem, where the convex core region is a complicated polytope.

In this backdrop, Gu and Effros estimated $|\Zcal_k|\le |\Xcal_k|$ for the Wyner-Ahlswede-K\"{o}rner problem using a linear programing argument \cite{GuEffros}. Later in \cite{GuJanaEffros}, the same result was extended to the Wyner-Ziv problem, and to the partial side information problem \cite{Upper}. The above result was crucially dependent on the fact that the convex core region that sweep out the overall inner bound is an orthant. In contrast, we shall prove the alphabet size $|\Zcal_k|\le |\Xcal_k|$ for any arbitrary problem, where the core region is always a polytope. Specifically, we decompose the polytope into constituent orthants, and make an orthant-based argument. The above decomposition, apart from being central to the problem at hand, enhances the geometric understanding of source coding. The main difficulty here lies in identifying the extreme points exhaustively, thereby identifying the constituent orthants. We show that there are $M!$ such orthants for an $M$-source problem. In order to prove this result, we develop an intricate chain of information theoretic results. Further, the orthant-based reasoning borrows an essential notion from a linear-programing-based argument. In particular, we consider only extreme points, which are reached by translating any hyperplane, with its direction fixed, away from the origin towards the achievable region. Our final argument about the alphabet size follows the line of Wyner and Ziv based on a version of Caratheodory's theorem \cite{WZ}.

\Section{Canonical Inner Bound}
\label{sec:CIB}

Consider joint source distribution $p(x_{\{1,...,M\}},s,v)$ governing source variables $X_{\{1,...,M\}}$, decoder side information $S$, and target variable $V$ for lossy reconstruction/estimation. Also consider $L$ bounded distortion measures $d_l: \Vcal\times \Vcalhat_l\rightarrow [0,\dlmax]$ ($1\le l\le L$), each with a possibly distinct reconstruction alphabet $\Vcalhat_l$. In this setting, the canonical inner bound $\Acal_1^*$ is defined as follows.

\begin{definition}
\label{def:A*} Define $\Acal_1^*$ as the set of $(M+L)$-vectors
$(R_{\{1,...,M\}},D_{\{1,...,L\}})$ satisfying the following conditions:
\begin{enumerate}
\item auxiliary random variables $Z_{\{1,...,M\}}$ (taking values in respective finite
alphabets $\Zcal_{\{1,...M\}}$) exist such that $Z_m=X_m$, $1\le m \le J$, and $(X_{\{1,...,M\}},S,V,Z_{\{J+1,...,M\}})$ follows the joint distribution
\be
\label{eq:p...}
p(x_{\{1,...,M\}},s,v)\prod_{k=J+1}^M q_k(z_k|x_k),\ee
for some test channels $\{q_k(z_k|x_k)\}_{k=J+1}^{M}$;
\item {\em (rate conditions)} \be \label{eq:R1}  I(X_I;Z_I|Z_{I^c},S)\le \sum_{i\in I}
R_i,
\ee where $I^c= \{1,2,...,M\}\setminus I$, and condition (\ref{eq:R1}) holds for all
$I\subseteq\{1,...,M\}\setminus \emptyset$;
\item {\em (distortion conditions)} mappings $\psi_l:\Xcal_1\times...
\times \Xcal_{J}\times\Zcal_{J+1}\times...\times\Zcal_M \times \Scal \rightarrow \Vcalhat_l$, $1\le l\le L$,
exist such that \be \label{eq:dist} \E
d_l(V,\psi_l(X_{\{1,...,J\}},Z_{\{J+1,...,M\}},S))\le D_l.
 \ee
 \end{enumerate}
\end{definition}

\begin{lemma}
\label{le:AlphSize}
Every extreme point of $\Acal_1^*$ corresponds to some choice of auxiliary variables $Z_{\{J+1,...,M\}}$ with alphabet sizes $|\Zcal_k|\le |\Xcal_k|$, $J+1\le k \le M$.
\end{lemma}

The main goal of this paper is to prove Lemma \ref{le:AlphSize}. The proof is difficult because
$\Acal_1^*$ has a complicated geometry. First of all, consider specific  auxiliary variables $Z_{\{J+1,...,M\}}$. Then choosing coordinate planes $y_i=R_i=0$, $1\le i \le M$, and $y_{M+l} = D_l=0$, $1\le l \le L$, note that distortion equations (\ref{eq:dist}) are all parallel to coordinate planes, and hence form an orthant, whose analysis is tractable. On the other hand, the rate equations (\ref{eq:R1}) are not all parallel to coordinate planes, leading to a complicated region. In this backdrop, in Sec.~\ref{sec:DARR} we consider the {\em distortion-extracted} rate region given by (\ref{eq:R1}), and find a decomposition into finite number of orthants. Based on such decomposition, in Sec.~\ref{sec:decompose} we write $\Acal_1^*$ as a finite union of component regions that are formed by orthants. Finally,  using such component regions, the extreme points in Lemma \ref{le:AlphSize} are characterized in Sec.~\ref{sec:linear} with the help of certain linear combination properties.

\Section{Geometry of Distortion-Extracted Rate Region}
\label{sec:DARR}

We first consider the rate region formed by rate conditions (\ref{eq:R1}). More generally, consider random variables $(X_{\{1,...,M\}},S,Z_{\{1,...,M\}})$ following the joint distribution
\be
\label{eq:p...!}
p(x_{\{1,...,M\}},s)\prod_{k=1}^M q_k(z_k|x_k).\ee
In this section, we fix $p(x_{\{1,...,M\}},s)$ as well as all $q_k(z_k|x_k)$, $1\le k \le M$. Further, define $\Bcal^*$ as the set of rate $M$-vectors
$R_{\{1,...,M\}}$ satisfying
\be \label{eq:R1'}
\label{eq:Rdef}
 I(X_I;Z_I|Z_{I^c},S)\le \sum_{i\in I}
R_i,
\ee where condition (\ref{eq:Rdef}) holds for all
$I\subseteq\{1,...,M\}\setminus \emptyset$.
We call $\Bcal^*$ the {\em distortion-extracted} rate region because it is delinked from distortion measures. Of course, we also do not impose the original restrictions $Z_m=X_m$, $1\le m \le J$. Next we find the extreme points of $\Bcal^*$. In our analysis, we shall assume that there is no degeneracy, i.e., any extraneous Markov chain property, not dictated by the form (\ref{eq:p...!}) of joint distribution $p$, does not hold. Note that the nondegeneracy requirement is mild, and met if all random variables under consideration are statistically dependent.

\subsection{Number of Extreme Points: Upper Bound}

\begin{lemma}
\label{le:unique}
Suppose there exists rate $M$-vector $R_{\{1,...,M\}}$ such that
\bea
\label{eq:I}
I\left(X_{I};Z_I|Z_{I^c},S\right) &=& \sum_{i\in I}
R_i\\
\label{eq:I'}
I\left(X_{I'};Z_{I'}|Z_{{I'}^c},S\right) &=& \sum_{i\in I'}
R_i
\eea
simultaneously hold for some distinct
sets $I,I'\subseteq \{1,...,M\}\setminus \emptyset$. Then either $I\subset I'$ or $I'\subset I$.
\end{lemma}

The proof is involved, and makes use of a series of new information-theoretic relations involving $(X_{\{1,...,M\}},S,Z_{\{1,...,M\}})$. It is given in Appendix
\ref{sec:ProofUnique}.

\begin{lemma}
\label{le:num} $\Bcal^*$ has at most $M!$ extreme points.
\end{lemma}

{\bf {\em Proof}:} At each extreme point of $\Bcal^*$, $M$ of the $2^M-1$ constraints given by (\ref{eq:R1'}) are active. Therefore, in view of Lemma \ref{le:unique}, the number of extreme points of $\Bcal^*$ is upper bounded by the number of possible ways one can have
$$ I^{(1)}\subset I^{(2)} \subset ... \subset I^{(m)} \subset I^{(m+1)}
\subset ... \subset I^{(M-1)}\subset I^{(M)},$$ where $I^{(m)} \subseteq \{1,...,M\}$ with cardinality $|I^{(m)}|=m$, $1\le m \le M$. To begin with, we have the only choice $I^{(M)} = \{1,...,M\}$. However, given any $I^{(m+1)}$ ($1\le m < M$), one can choose $I^{(m)}$ by discarding one of the $m+1$ elements of $I^{(m+1)}$. Hence one can choose the entire sequence of sets $\{I^{(m)}\}_{m=1}^{M}$ in $M\times (M-1) \times ...\times 2 = M!$ possible ways. Hence the result. \hfill$\Box$

\begin{remark}
The above argument does not clarify whether all $M!$ points under consideration are distinct. Hence we can claim only an upper bound.
\end{remark}

\subsection{Number of Extreme Points: Lower Bound}

\begin{lemma}
\label{le:cornQ}
The rate $M$-vector $R_{\{1,...,M\}}$ such that
\be
\label{eq:cornQ}
R_{i} = I(X_{i};Z_{i}|Z_{\{1,...,i-1\}},S), \quad 1\le i\le M,
\ee
is an extreme point of $\Bcal^*$.
\end{lemma}

\begin{remark}
By Lemma \ref{le:last} and (\ref{eq:cornQ}), we have
\be
\label{eq:lastL}
I\left(X_{I};Z_{I}|Z_{I^c},S\right) \le
\sum_{i\in I} I\left(X_i;Z_i|Z_{\{1,...,i-1\}},S\right) = \sum_{i\in I} R_i
\ee
for all $I\subseteq \{1,...,M\}\setminus\emptyset$. Thus, by (\ref{eq:Rdef}), $R_{\{1,...,M\}}\in \Bcal^*$.
\end{remark}

{\bf {\em Proof}:} It is enough to show that the given $R_{\{1,...,M\}}$ makes $M$ constraints, given in (\ref{eq:Rdef}), active. From (\ref{eq:cornQ}), we can write
\be
\label{eq:cornQ1}
\sum_{i=m}^{M} I(X_{i};Z_{i}|Z_{I_{\{1,...,i-1\}}},S) = \sum_{i=m}^{M} R_i
\ee
for each $1\le m\le M$.
Further, by Corollary \ref{cor:chain3}, (\ref{eq:cornQ1}) is same as
\be
\label{eq:cc3'}
I\left(X_{\{m,...,M\}};Z_{\{m,...,M\}}|
Z_{\{1,...,m-1\}},S\right)= \sum_{i=m}^{M} R_i, \quad 1\le m \le M,
\ee
which makes $M$ constraints, given in (\ref{eq:Rdef}), active.
This completes the proof. \hfill$\Box$

Now the indices $\{1,...,M\}$ in (\ref{eq:cornQ}) can be permuted to obtain $M!$ extreme points. Importantly, these extreme points are all distinct due to the nondegeneracy assumption.

\begin{corollary}
\label{cor:lower} $\Bcal^*$ has at least $M!$ extreme points.
\end{corollary}

\begin{remark}
\label{rem:num}
By Lemma \ref{le:num} and Corollary \ref{cor:lower}, $\Bcal^*$ has exactly $M!$ extreme points, each of which takes the form (\ref{eq:cornQ}) except that the indices $\{1,...,M\}$ undergo suitable permutation. (As it is, (\ref{eq:cornQ}) corresponds to identity permutation.)
\end{remark}

\Section{Decomposition of \boldmath{$\Acal_1^*$}}
\label{sec:decompose}

Now we move on to the rate-distortion region $\Acal_1^*$.
Specifically, consider subset $\Acal_1^*(\{q_k\})$ of $\Acal_1^*$ defined by (\ref{eq:p...})--(\ref{eq:dist}) for given conditional distributions $q_k(z_k|x_k)$, $J+1\le k \le M$. Of course, $\Acal_1^* = \bigcup \Acal_1^*(\{q_k\})$, where the union is taken over all $\{q_k\}$. Note that, like $\Acal_1^*$, $\Acal_1^*(\{q_k\})$ is a subset of the $(M+L)$-dimensional real space. However, although $\Acal_1^*$ is not necessarily convex, each $\Acal_1^*(\{q_k\})$ is convex. Further, every extreme point of $\Acal_1^*$ is an extreme point of some $\Acal_1^*(\{q_k\})$.  Finally, notice that the projection of $\Acal_1^*(\{q_k\})$ onto the space of $M$ rate coordinates is the same as $\Bcal^*$ with the choice $Z_m=X_m$, $1\le m\le J$ (which does not violate our degeneracy assumption), whereas the projection onto the space of $L$ distortion coordinates is simply a suitable orthant. Therefore, by Remark \ref{rem:num}, $\Acal_1^*(\{q_k\})$ possesses $M!$ extreme points, one of which, denoted $(R^0_{\{1,...,M\}}(\{q_k\}),D^0_{\{1,...,L\}}(\{q_k\}))$, is specified by (from (\ref{eq:cornQ}) and (\ref{eq:dist}))
\bea
\label{eq:cornQ'}
R^0_{i}(\{q_k\}) &=& I(X_{i};Z_{i}|Z_{\{1,...,i-1\}},S), \quad 1\le i\le M\\
\label{eq:dist'}
D^0_l(\{q_k\}) &=&
\min_{\psi_l} \E
d_l(V,\psi_l(Z_{\{1,...,M\}},S)), \quad 1\le l\le L,
\eea
where $Z_m=X_m$, $1\le m\le J$. In general, any extreme point $(R^\pi_{\{1,...,M\}}(\{q_k\}),D^\pi_{\{1,...,L\}}(\{q_k\}))$ is
generated by a suitable permutation (bijection) $P^\pi: \{1,...,M\} \rightarrow \{1,...,M\}$, where $\pi$ takes $M!$ values, say, $\{0,...,M!-1\}$ (we set $P^0$ to be the identity permutation).
In other words, in (\ref{eq:cornQ'}) and (\ref{eq:dist'}), each occurrence of index $i$ is replaced by $P_\pi(i)$.
As regards dependence on $\pi$, vectors $R^\pi_{\{1,...,M\}}(\{q_k\})$ are all distinct (as mentioned earlier), whereas vectors $D^\pi_{\{1,...,L\}}(\{q_k\})$ are all identical.

At this point, denote the orthant specified by $(R^\pi_{\{1,...,M\}}(\{q_k\}),D^\pi_{\{1,...,L\}}(\{q_k\}))$ as
\be
\label{eq:Orth0}
\Acal_{1;\pi}^*(\{q_k\}) = \{(R_{\{1,...,M\}},D_{\{1,...,L\}}):R^\pi_{i}(\{q_k\}) \le R_i,1\le i\le M; D^\pi_{l}(\{q_k\}) \le D_l,1\le l\le L\}
\ee
for $0\le \pi \le M!-1$, and all possible $\{q_k\}$.
Clearly,
$$\Acal_1^*(\{q_k\}) = \conv\left(\bigcup_{\pi=0}^{M!-1} \Acal_{1;\pi}^*(\{q_k\})\right),$$
where $\conv(\cdot)$ indicates `convex hull of'. Consequently, we have
\be
\label{eq:Orth1}
\conv(\Acal_1^*) = \conv\left(\bigcup_{\{q_k\}} \Acal_1^*(\{q_k\})\right) = \conv\left(\bigcup_{\{q_k\}} \bigcup_{\pi=0}^{M!-1} \Acal_{1;\pi}^*(\{q_k\})\right).
\ee
Now, interchanging the union operations in the last term in (\ref{eq:Orth1}), and defining
\be
\label{eq:A1pi*}
\Acal_{1;\pi}^*=  \bigcup_{\{q_k\}}  \Acal_{1;\pi}^*(\{q_k\}),
\ee
we obtain
\be
\label{eq:Orth2}
\conv(\Acal_1^*) = \conv\left( \bigcup_{\pi=0}^{M!-1} \Acal_{1;\pi}^*\right).
\ee
In view of (\ref{eq:Orth2}), every extreme point of $\Acal_1^*$ is an extreme point of some $\Acal_{1;\pi}^*$. Consequently, in order to establish Lemma \ref{le:AlphSize}, it is enough to show the following.

\begin{lemma}
\label{le:AlphSizePI}
Every extreme point of $\Acal_{1;\pi}^*$ ($0\le \pi \le M!-1$) corresponds to some choice of auxiliary variables $Z_{\{J+1,...,M\}}$ with alphabet sizes $|\Zcal_k|\le |\Xcal_k|$, $J+1\le k \le M$.
\end{lemma}

The rest of the note is devoted to the proof of Lemma \ref{le:AlphSizePI}. In particular, we shall prove the result only for $\pi=0$. Our analysis extends to other values of $\pi$ in a straightforward manner. At present, consider the real $(M+L)$-space, and let $y_i=0$, $1\le i \le M+L$, be the coordinate planes. In this space, an $(M+L-1)$-hyperplane
\be
\label{eq:HP}
\sum_{i=1}^{M+L} a_i y_i = c
\ee
is specified by the direction cosine vector $(a_1,...,a_{M+L})$ subject to $\sum_{i=1}^{M+L} a_i^2 = 1$, and the intercept $c$. At this point, identifying $y_i=R_i$, $1\le i \le M$, and $y_{M+l}=D_l$, $1\le l \le L$, note that $\Acal_{1;0}^*$ lies in the nonnegative orthant. Further, every extreme point of $\Acal_{1;0}^*$ has a tangent hyperplane of the form (\ref{eq:HP}), whose direction cosines and intercept are nonnegative ($a_i\ge 0$, $1\le i \le M+L$; $c\ge0$). Conversely, for any $(a_1,...,a_{M+L})$ with $a_i\ge 0$, $1\le i \le M+L$, there exists $c\ge 0$ such that the hyperplane (\ref{eq:HP}) is tangent to $\Acal_{1;0}^*$ at some extreme point. Hence we obtain the following result.

\begin{corollary}
\label{cor:exA10}
The set
%$\Ecal_{1;0}^*$
of extreme points of $\Acal_{1;0}^*$ is given by
\be
\label{eq:extreme..}
\left\{\arg \min_{(R_{\{1,...,M\}},D_{\{1,...,L\}})\in\Acal_{1;0}^*} \left(\sum_{i=1}^M a_i R_i + \sum_{l=1}^L a_{M+l} D_l \right): \sum_{i=1}^{M+L} a_i^2 = 1; a_i \ge 0,1\le i \le M+L\right\}.
\ee
\end{corollary}

By (\ref{eq:Orth0}) and (\ref{eq:A1pi*}), every minimizer in (\ref{eq:extreme..}) is of the form $(R^0_{\{1,...,M\}}(\{q_k\}),D^0_{\{1,...,L\}}(\{q_k\}))$ for some $\{q_k\}$. Further, using $Z_m=X_m$, $1\le m\le J$, in (\ref{eq:cornQ'}), notice that $R^0_{\{1,...,J\}}(\{q_k\})$  does not depend on $(\{q_k\})$. Hence we set $a_1=...=a_J=0$ without loss of generality (and scale the remaining direction cosines appropriately) to obtain the following.

\begin{corollary}
\label{cor:exA10'}
The set
%$\Ecal_{1;0}^*$
of extreme points of $\Acal_{1;0}^*$ is given by the set of rate-distortion vectors $(R^0_{\{1,...,M\}}(\{q_k\}),D^0_{\{1,...,L\}}(\{q_k\}))$ such that $\{q_k\}$ minimizes
\be
\label{eq:exA10'}
\sum_{i=J+1}^M a_i R^0_i(\{q_k\}) + \sum_{l=1}^L a_{M+l} D_l^0(\{q_k\}),
\ee
and direction cosine vector $a_{\{J+1,...,M+L\}}$ varies through admissible values.
\end{corollary}

Note that Lemma \ref{le:AlphSizePI} follows for $\pi =0$ (corresponding to identity permutation $P^0$), if we lose no generality by restricting to minimizers $\{q_k\}$ of (\ref{eq:exA10'}) that satisfy $|\Zcal_k|\le |\Xcal_k|$, $J+1\le k \le M$. We shall show that the last condition indeed holds as a consequence of certain linear combination properties.

\Section{Linear Combination Properties}
\label{sec:linear}

\subsection{Change of Variables}

For $J+1\le k\le M$, denote marginal distributions of $X_k$ and $Z_k$ by $p_k(x_k)$ and $p'_k(z_k)$, respectively, and conditional distribution of $X_k$ given $Z_k$ by $q'_k(x_k|z_k)$. Note that $p_k(x_k)$ is specified by marginalizing the source distribution $p(x_{\{1,...,M\}},s,v)$. Further, by Bayes' rule, we have $p_k(x_k) q_k(z_k|x_k) = p'_k(z_k)q'_k(x_k|z_k)$. Of course, one completely specifies both $p'_k$ and $q'_k$ by specifying $q_k$. At the same time, rather than varying $q_k$, we can equivalently vary the pair $(p'_k,q'_k)$ subject to the admissibility condition
\be
\label{eq:pCON}
p_k(x_k) = \sum_{z_k\in\Zcal_k} p'_k(z_k)q'_k(x_k|z_k).
\ee
Apart from the above specific notation, we shall denote by `$r$' generic distributions. For example, $r(y,u|w)$ indicates the joint distribution of $(Y,U)$ conditioned on $W$.

At this point, consider identity permutation $P^0$ of $\{1,...,M\}$, and, correspondingly, the set $\Acal_{1;0}^*(\{p'_k,q'_k\})$. Here, we recall that variation of $\{q_k\}$, and variation of $\{p'_k,q'_k\}$ subject to (\ref{eq:pCON}) are equivalent, and,
in a slight abuse of notation, denote by $\Acal_{1;0}^*(\{p'_k,q'_k\})$ the set function of $\{p'_k,q'_k\}$ equalling $\Acal_{1;\pi}^*(\{q_k\})$. Subsequently, we shall make analogous change of variables without explicit mention. Using $Z_m=X_m$, $1\le m\le J$, in (\ref{eq:cornQ'}) and (\ref{eq:dist'}), we have
\bea
\label{eq:cornQ''1}
R^0_{i}(\{p'_k,q'_k\}) &=& H(X_{i}|X_{\{1,...,i-1\}},S), \quad 1\le i\le J\\
\label{eq:cornQ''2}
R^0_{i}(\{p'_k,q'_k\}) &=& I(X_{i};Z_i|X_{\{1,...,J\}},Z_{\{J+1,...,i-1\}},S), \quad J+1\le i\le M\\
\label{eq:dist''}
D^0_l(\{p'_k,q'_k\}) &=&
\min_{\psi_l} \E
d_l(V,\psi_l(X_{\{1,...,J\}},Z_{\{J+1,...,M\}},S)), \quad 1\le l\le L.
\eea
As mentioned earlier, and by (\ref{eq:cornQ''1}), $R^0_{\{1,...,J\}}(\{p'_k,q'_k\})$ does not depend on $(\{p'_k,q'_k\})$. However, the remaining rate and distortion components, given by (\ref{eq:cornQ''2}) and (\ref{eq:dist''}), do exhibit dependence on $(\{p'_k,q'_k\})$.

Next we isolate the dependence of individual rate as well as distortion component on individual pair $(p'_k,q'_k)$, while keeping the rest of the pairs fixed. We highlight the dependence on $(p'_k,q'_k)$ by dropping the rest of the pairs $\{(p'_\kappa,q'_\kappa)\}_{\kappa \ne k}$ from the argument. Specifically, we show that each $R_i^0(p'_k,q'_k)$ ($k\le i\le M$) and each $D_l^0(p'_k,q'_k)$ ($1\le l\le L$) is a linear combination of functionals of $q'_k(\cdot|z_k)$'s weighted by $p'_k(z_k)$'s. Here $q'_k(\cdot|z_k)$ denotes the probability vector $\{q'_k(x_k|z_k)\}_{x_k\in\Xcal_k}$ for a given $z_k\in \Zcal_k$.

\subsection{Rate Components}

Consider $J+1\le k\le i\le M$. From (\ref{eq:cornQ''2}), we have
\bea
\nonumber
R^0_{i}(p'_k,q'_k) &=& I(X_{i};Z_i|X_{\{1,...,J\}},Z_{\{J+1,...,i-1\}},S)\\
\label{eq:IH1}
&=& H(X_i|X_{\{1,...,J\}},Z_{\{J+1,...,i-1\}},S) - H(X_i|X_{\{1,...,J\}},Z_{\{J+1,...,i\}},S).
\eea
Further, denote by $\Delta_{\Xcal_k}$ the $(|\Xcal_k|-1)$-dimensional probability simplex, i.e.,
the set of probability vectors defined on $\Xcal_k$.

\begin{lemma}
\label{le:RateLin1}
If $J+1\le k \le i \le M$, then $$H(X_i|X_{\{1,...,J\}},Z_{\{J+1,...,i-1\}},S) = \sum_{z_k\in\Zcal_k} p'_k(z_k) \Phi^{(1)}_{ki}(q'_k(\cdot|z_k))$$
for some functional $\Phi^{(1)}_{ki}$ defined on $\Delta_{\Xcal_k}$.
\end{lemma}

{\bf {\em Proof}:} First note that, if $i=k$, then the target entropy does not depend on $(p'_k,q'_k)$, and $\Phi^{(1)}_{ki}$ reduces to a trivial constant. A more interesting situation arises when $i > k$. In this case, verify that $k\in\{J+1,...,i-1\}$. Now write $U = (X_{\{1,...,J\}},Z_{\{J+1,...,i-1\}\setminus\{k\}},S)$, and verify that
\be
\label{eq:Mlop0}
Z_k \rightarrow X_k \rightarrow (U,X_i)
\ee
forms a Markov chain. Hence we obtain
\bea
\nonumber
&&H(X_i|X_{\{1,...,J\}},Z_{\{J+1,...,i-1\}},S)
~~=~~ H(X_i|Z_k,U)\\
\nonumber
&=& -\sum_{(x_i,z_k,u)} r(x_i,z_k,u)\log\frac{r(x_i,z_k,u)}{r(z_k,u)}\\
\label{eq:MLop1}
&=& -\sum_{(x_i,z_k,u)} \sum_{x_k} p'_k(z_k)q'_k(x_k|z_k)r(x_i,u|x_k)\log\frac{\sum_{x_k} p'_k(z_k)q'_k(x_k|z_k)r(x_i,u|x_k)}{\sum_{x_k} p'_k(z_k)q'_k(x_k|z_k)r(u|x_k)}\\
\label{eq:MLop2}
&=& -\sum_{z_k} p'_k(z_k) \sum_{(x_i,u)} \sum_{x_k}
q'_k(x_k|z_k)r(x_i,u|x_k)\log\frac{\sum_{x_k} q'_k(x_k|z_k)r(x_i,u|x_k)}{\sum_{x_k} q'_k(x_k|z_k)r(u|x_k)}\\
\label{eq:MLop3}
&=& \sum_{z_k} p'_k(z_k) \Phi^{(1)}_{ki}(q'_k(\cdot|z_k)).
\eea
Here (\ref{eq:MLop1}) follows by noting Markov chain (\ref{eq:Mlop0}), and writing
\beann
r(z_k,x_i,u) &=& \sum_{x_k} r(z_k, x_k,x_i,u)~=~ \sum_{x_k} p'_k(z_k)q'_k(x_k|z_k)r(x_i,u|x_k)\\
r(z_k,u) &=& \sum_{x_k} p'_k(z_k)q'_k(x_k|z_k)r(u|x_k).
\eeann
Further, (\ref{eq:MLop2}) follows by rearranging, and by canceling out $p'_k(z_k)$ from the numerator and denominator of the argument of `$\log$'. Finally, (\ref{eq:MLop3}) follows by defining the functional $$\Phi^{(1)}_{ki}(t) = -\sum_{(x_i,u)} \sum_{x_k}
t(x_k) r(x_i,u|x_k)\log\frac{\sum_{x_k} t(x_k) r(x_i,u|x_k)}{\sum_{x_k} t(x_k) r(u|x_k)},$$
where $t=\{t(x_k):x_k\in \Xcal_k\}$ is any probability vector on $\Xcal_k$. \hfill $\Box$

Adopting a similar approach, we also obtain the following.

\begin{lemma}
\label{le:RateLin2}
If $J+1\le k \le i \le M$, then $$H(X_i|X_{\{1,...,J\}},Z_{\{J+1,...,i\}},S) = \sum_{z_k\in\Zcal_k} p'_k(z_k) \Phi^{(2)}_{ki}(q'_k(\cdot|z_k))$$
for some functional $\Phi^{(2)}_{ki}$ defined on $\Delta_{\Xcal_k}$.
\end{lemma}

Noting (\ref{eq:IH1}),
combining Lemmas \ref{le:RateLin1} and \ref{le:RateLin2},
and writing $\Phi_{ki} = \Phi^{(1)}_{ki}-\Phi^{(2)}_{ki}$, we obtain the following corollary.

\begin{corollary}
\label{cor:RateLin}
If $J+1\le k \le i \le M$, then $$R^0_i(p'_k,q'_k) = \sum_{z_k\in\Zcal_k} p'_k(z_k) \Phi_{ki}(q'_k(\cdot|z_k))$$
for some functional $\Phi_{ki}$ defined on $\Delta_{\Xcal_k}$.
\end{corollary}

\subsection{Distortion Components}

\begin{lemma}
\label{le:DistLin}
 For $J+1\le k \le M$, and $1\le l\le L$, we have $$D^0_l(p'_k,q'_k) = \sum_{z_k\in\Zcal_k} p'_k(z_k) \Psi_{kl}(q'_k(\cdot|z_k))$$
for some functional $\Psi_{kl}$ defined on $\Delta_{\Xcal_k}$.
\end{lemma}

{\bf {\em Proof}:}
Write $U = (X_{\{1,...,J\}},Z_{\{J+1,...,M\}\setminus\{k\}},S)$, and verify that
\be
\label{eq:Kab0}
Z_k \rightarrow X_k \rightarrow (U,V)
\ee
forms a Markov chain. Hence from (\ref{eq:dist''}), we obtain
\bea
\nonumber
D^0_l(p'_k,q'_k) &=&
\min_{\psi_l} \E
d_l(V,\psi_l(U,Z_k))\\
\nonumber
&=& \min_{\psi_l} \sum_{(v,u,z_k)} r(u,v,z_k) d_l(v,\psi_l(u,z_k))\\
\label{eq:Kab1}
&=& \min_{\psi_l} \sum_{(v,u,z_k)} \sum_{x_k} p'_k(z_k) q'_k(x_k|z_k) r(u,v|x_k) d_l(v,\psi_l(u,z_k))\\
\label{eq:Kab2}
&=&
\sum_{z_k} p'_k(z_k)
\sum_{u} \min_{\vhat_l} \left[
\sum_{(v,x_k)} q'_k(x_k|z_k) r(u,v|x_k) d_l(v,\vhat_l)\right]\\
\label{eq:Kab3}
&=& \sum_{z_k} p'_k(z_k) \Psi_{kl}(q'_k(\cdot|z_k)).
\eea
Here (\ref{eq:Kab1}) follows by noting Markov chain (\ref{eq:Kab0}), and writing
$$
r(u,v,z_k) = \sum_{x_k} r(u,v,x_k,z_k) = \sum_{x_k} p'_k(z_k)q'_k(x_k|z_k)r(u,v|x_k).$$
Further, (\ref{eq:Kab2}) follows by rearranging. Finally, (\ref{eq:Kab3}) follows by defining the functional $$\Psi_{kl}(t) = \sum_{u} \min_{\vhat_l} \left[
\sum_{(v,x_k)} t(x_k) r(u,v|x_k) d_l(v,\vhat_l)\right],$$
where $t=\{t(x_k):x_k\in \Xcal_k\}$ is any probability vector on $\Xcal_k$. \hfill $\Box$

\subsection{Minimization of Linear Combination}

At this time, consider the setting of Corollary
\ref{cor:exA10'}, i.e., $a_1=...=a_J=0$.

\begin{lemma}
\label{le:Linear}
Pick any $J+1\le k\le M$, and fix admissible $a_{\{J+1,...,M+L\}}$ and $\{(p'_\kappa,q'_\kappa)\}_{\kappa\ne k}$ in an arbitrary manner. Then there exists a minimizer $(p'_k,q'_k)$ of the problem $$\min_{\mbox{\footnotesize $(p'_k,q'_k)$ subject to (\ref{eq:pCON})}} \sum_{i=J+1}^M a_i R_i^0(p'_k,q'_k) + \sum_{l=1}^L a_{M+l} D_l^0(p'_k,q'_k)$$ such that $p'_k(z_k)$ is defined on alphabet $\Zcal_k$ with size $|\Zcal_k|\le |\Xcal_k|$ (and hence $q'_k(x_k|z_k)$ is specified by at most $|\Xcal_k|$ probability vectors defined on $\Xcal_k$).
\end{lemma}

{\bf {\em Proof}:} Given $a_{\{J+1,...,M+L\}}$ and $\{(p'_\kappa,q'_\kappa)\}_{\kappa\ne k}$, consider $$\omega = \sum_{i=J+1}^M a_i R_i^0(p'_k,q'_k) + \sum_{l=1}^L a_{M+l} D_l^0(p'_k,q'_k),$$ and
denote by $\Omega$ the set of admissible values of $\omega$. Further, denote $\omega^* = \min_{\omega\in \Omega} \omega$.
Now,
by Corollary \ref{cor:RateLin} and Lemma \ref{le:DistLin}, we have
\be
\label{eq:lin57}
\omega = \sum_{z_k\in\Zcal_k} p'_k(z_k) \Theta(q'_k(\cdot|z_k)),
\ee
where $\Theta(t) = \sum_{i=J+1}^M a_i \Phi_{ki}(t)+ \sum_{l=1}^L a_{M+l} \Psi_{kl}(t)$ is defined on $\Delta_{\Xcal_k}$. Note that $\Theta$ is continuous and bounded, and the $(|\Xcal_k|-1)$-dimensional probability simplex $\Delta_{\Xcal_k}$ is compact. Now consider the mapping $t\rightarrow(t,\Theta(t))$, and denote by $\Scal$ the image of  $\Delta_{\Xcal_k}$ under this mapping. Of course, $\Scal$ is connected and compact, and $\Scal$ has dimensionality $|\Xcal_k|$. Therefore, by Fenchel-Eggleston strengthening of Caratheodory's theorem, any point in $\conv(\Scal)$ is a linear combination of at most $|\Xcal_k|$ points in $\Scal$.
Further, in view of (\ref{eq:pCON}) and (\ref{eq:lin57}), any pair $(p_k,\omega)$ belongs to $\conv(\Scal)$. In particular, set $\Omega$ of admissible $\omega$, where source distribution $p_k$ is fixed by problem statement, is given by
$$\Omega = \{\omega:(p_k,\omega)\in \conv(\Scal)\}.$$
In other words, every admissible $\omega\in \Omega$, including $\omega^*$, can be expressed as in (\ref{eq:lin57}) with $|\Zcal_k|\le |\Xcal_k|$. This completes the proof. \hfill $\Box$

\begin{corollary}
\label{cor:Linear}
For any admissible $a_{\{J+1,...,M+L\}}$, there exists a minimizer $\{p'_k,q'_k\}$ of the problem $$\min_{\mbox{\footnotesize $\{p'_k,q'_k\}$ subject to (\ref{eq:pCON})}} \sum_{i=J+1}^M a_i R_i^0(\{p'_k,q'_k\}) + \sum_{l=1}^L a_{M+l} D_l^0(\{p'_k,q'_k\})$$ such that each $p'_k(z_k)$ ($J+1\le k\le M$) is defined on alphabet $\Zcal_k$ with size $|\Zcal_k|\le |\Xcal_k|$ (and hence each $q'_k(x_k|z_k)$ is specified by at most $|\Xcal_k|$ probability vectors defined on $\Xcal_k$).
\end{corollary}

{\bf {\em Proof}:} We shall prove the result by contradiction. Suppose there exists admissible $a_{\{J+1,...,M+L\}}$ such that a minimizer $\{p'_k,q'_k\}$ with $|\Zcal_k|\le |\Xcal_k|$, $J+1\le k\le M$, does not exist. Pick such $a_{\{J+1,...,M+L\}}$, and compute the minimum value $\phi$ of the objective function. By supposition, any corresponding minimizer $\{p'_k,q'_k\}$ has $|\Zcal_i| > |\Xcal_i|$ for some $J+1\le i\le M$. We now undertake a procedure such that the minimum value does not increase at any stage. Specifically,
choose $k=J+1$, and keep $\{(p'_\kappa,q'_\kappa)\}_{\kappa \ne k}$ fixed. By Lemma \ref{le:Linear}, the objective function is no greater than $\phi$ for some new choice $(p'_k,q'_k)$ with $|\Zcal_k|\le |\Xcal_k|$. Update $(p'_k,q'_k)$ to this new choice. Next choose $k=J+2$, keep $\{(p'_\kappa,q'_\kappa)\}_{\kappa \ne k}$ fixed, and update $(p'_k,q'_k)$ (in view of Lemma \ref{le:Linear}) such that the objective function is no greater than $\phi$, yet $|\Zcal_k|\le |\Xcal_k|$. Continue this procedure till $k=M$. Finally, we have a new $\{(p'_k,q'_k)\}$ with $|\Zcal_k|\le |\Xcal_k|$, $J+1\le k\le M$, such that the corresponding objective function is no greater than $\phi$. This is a contradiction. \hfill$\Box$

{\bf {\em Proofs of Lemmas \ref{le:AlphSizePI} and \ref{le:AlphSize}}:} Note that $\{q_k\}$ is completely determined by $\{p'_k,q'_k\}$ by Bayes' rule $$q_k(z_k|x_k) = p'_k(z_k)q'_k(x_k|z_k)/p_k(x_k),$$ because $p_k(x_k)$ is specified by the problem statement. Therefore, by Corollary \ref{cor:Linear}, we lose no generality by restricting to minimizers $\{q_k\}$ of (\ref{eq:exA10'}) that satisfy $|\Zcal_k|\le |\Xcal_k|$, $J+1\le k \le M$. Hence Lemma \ref{le:AlphSizePI} follows for $\pi =0$ (corresponding to identity permutation $P^0$). Further, a similar analysis straightforwardly establishes Lemma \ref{le:AlphSizePI} for each $1\le \pi \le M!-1$. Finally, in view of (\ref{eq:Orth2}), Lemma \ref{le:AlphSize} follows. \hfill$\Box$

%\Section{Extreme Point and Lagrangian Minimization}
%\label{sec:lagrange}
%\input{lagrange}

\appendix

\setcounter{equation}{0}
\setcounter{figure}{0}
\setcounter{table}{0}
\renewcommand{\theequation}{\Alph{section}.\arabic{equation}}
\renewcommand{\thetable}{\Alph{section}.\arabic{table}}
\renewcommand{\thefigure}{\Alph{section}.\arabic{figure}}

\Section{Proof of Lemma \ref{le:unique}}
\label{sec:ProofUnique}

\begin{lemma}
\label{le:chain'}
Suppose sets $I,I'\subseteq \{1,...,M\}\setminus \emptyset$ are disjoint. Then
\be
\label{eq:rule'}
I\left(X_{I};Z_{I}|Z_{{(I\cup I')}^c},S\right)
= I\left(X_{I};Z_{I}|Z_{{I}^c},S\right)+ I\left(Z_I;Z_{I'}|Z_{{(I\cup I')}^c},S\right).
\ee
\end{lemma}

{\bf {\em Proof}:} First expand
\be
\label{eq:ami1}
I\left(Z_I;X_{I},Z_{I'}|Z_{{(I\cup I')}^c},S\right)
= I\left(Z_I;Z_{I'}|Z_{{(I\cup I')}^c},S\right) + I\left(Z_{I};X_{I}|Z_{{I}^c},S\right),
\ee
applying the chain rule of mutual entropy. Expand the same quantity again, now applying the chain rule in a different order:
\be
\label{eq:ami2.m}
I\left(Z_I;X_{I},Z_{I'}|Z_{{(I\cup I')}^c},S\right)
= I\left(Z_{I};X_{I}|Z_{{(I\cup I')}^c},S\right)
+ I\left(Z_I;Z_{I'}|X_{I},Z_{{(I\cup I')}^c},S\right).
\ee
Note that $Z_I\rightarrow (X_{I},Z_{{(I\cup I')}^c},S) \rightarrow
Z_{I'}$ forms a Markov chain (since $I$ and $I'$ are distinct), i.e.,
$$I\left(Z_I;Z_{I'}|X_{I},Z_{{(I\cup I')}^c},S\right)
=0$$ in (\ref{eq:ami2.m}). Hence, equating the right-hand sides of (\ref{eq:ami1})
and (\ref{eq:ami2.m}),
and rearranging, we obtain (\ref{eq:rule'}). \hfill$\Box$

\begin{lemma}
\label{le:chain}
Suppose sets $I,I'\subseteq \{1,...,M\}\setminus \emptyset$ are disjoint. Then
\bea
\label{eq:rule1}
I\left(X_{I\cup I'};Z_{I\cup I'}|Z_{{(I\cup I')}^c},S\right)
&=& I\left(X_{I};Z_{I}|Z_{{(I\cup I')}^c},S\right)
+
I\left(X_{I'};Z_{I'}|Z_{{I'}^c},S\right).
\eea
\end{lemma}

{\bf {\em Proof}:} For any quadruple $(U_1,U_2,V_1,V_2)$ of random variables, we can write
\bea
\nonumber
I(U_1,U_2;V_1,V_2) &=& I(U_1,U_2;V_1) + I(U_1,U_2;V_2|V_1)\\
\label{eq:red1}
&=&
I(U_1;V_1) + I(U_2;V_1|U_1) + I(U_2;V_2|V_1) + I(U_1;V_2|V_1,U_2)
\eea
by repeatedly applying the chain rule of mutual information.
Identifying $(U_1,U_2,V_1,V_2)$ with $(X_{I},X_{I'},Z_{I},Z_{I'})$, and applying formula (\ref{eq:red1}) (while maintaining conditioning on $(Z_{{(I\cup I')}^c},S)$ throughout), we obtain
\bea
\nonumber
I\left(X_{I\cup I'};Z_{I\cup I'}|Z_{{(I\cup I')}^c},S\right)
&=&
I\left(X_{I};Z_{I}|Z_{{(I\cup I')}^c},S\right)
+
I\left(X_{I'};Z_{I}|X_I,Z_{{(I\cup I')}^c},S\right)\\
\label{eq:red2}
&&\!\!\! +~
I\left(X_{I'};Z_{I'}|Z_{{I'}^c},S\right)
+
I\left(X_{I};Z_{I'}|X_{I'},Z_{{I'}^c},S\right).
\eea
In (\ref{eq:red2}),
$I\left(X_{I'};Z_{I}|X_I,Z_{{(I\cup I')}^c},S\right) =0$ and
$I\left(X_{I};Z_{I'}|X_{I'},Z_{{I'}^c},S\right) =0$, respectively, because $Z_{I} \rightarrow (X_I,Z_{{(I\cup I')}^c},S) \rightarrow X_{I'}$ and $Z_{I'} \rightarrow (X_{I'},Z_{{I'}^c},S) \rightarrow X_{I}$ form Markov chains (since $I$ and $I'$ are distinct). Hence the result.
\hfill$\Box$

More generally, any $\Ihat \subseteq \{1,...,M\}\setminus \emptyset$ can play the role of $\{1,...,M\}$ in the statement of Lemma \ref{le:chain} so that ${I'}^c$ can be replaced by $\Ihat\setminus I'$ and ${(I\cup I')}^c$ by $\Ihat \setminus ({I\cup I'})$. In that case, Lemma \ref{le:chain} immediately takes the following form:

\begin{corollary}
\label{cor:chain1}
Suppose sets $I,I'\subseteq \Ihat $ are disjoint, where $\Ihat \subseteq \{1,...,M\}\setminus \emptyset$.
Then
\bea
\label{eq:rule1'}
\!\!\!\!\!\!
\!\!\!\!\!\!\!\!\!\!\!\!
I\left(X_{I\cup I'};Z_{I\cup I'}|Z_{{\Ihat \setminus(I\cup I')}},S\right)
&=& I\left(X_{I};Z_{I}|Z_{{\Ihat \setminus(I\cup I')}},S\right)
+
I\left(X_{I'};Z_{I'}|Z_{{\Ihat \setminus I'}},S\right).
\eea
\end{corollary}

Now consider arbitrary $I\subseteq \{1,...,M\}\setminus \emptyset$ with cardinality $|I|=m$, and denote its elements by $i(1:m)$. Further, setting $\Ihat=\{1,...,M\}$, and letting $(\{i(1)\}, I\setminus\{i(1)\})$ play the role of $(I,I')$
in (\ref{eq:rule1'}), we have
\bea
\label{eq:eka1}
\!\!\!\!\!\!
\!\!\!\!\!\!
I\left(X_{I};Z_{I}|Z_{I^c},S\right)
&=& I\left(X_{i(1)};Z_{i(1)}|Z_{I^c},S\right)
+
I\left(X_{I\setminus\{i(1)\}};Z_{I\setminus\{i(1)\}}|
Z_{{(I\setminus\{i(1)\})}^c},S\right).\quad
\eea
Next set $\Ihat = I\setminus\{i(1)\} = \{i(2:m)\}$, let
$(\{i(2)\}, I\setminus\{i(1:2)\})$ play the role of $(I,I')$
in (\ref{eq:rule1'}), and continue so as to obtain
\bea
\label{eq:eka2}
I\left(X_{I};Z_{I}|
Z_{I^c},S\right)
&=& \sum_{j=1}^m
I\left(X_{i(j)};Z_{i(j)}|Z_{{(I\setminus \{i(1:j-1)\})}^c},S\right).
\eea
Noting ${(I\setminus \{i(1:j-1)\})}^c = \{1,...,M\}\setminus \{i(j:m)\}$ in (\ref{eq:eka2}), we have the following result.

\begin{corollary}
\label{cor:mid}
Suppose set $I\subseteq \{1,...,M\}\setminus \emptyset$ has cardinality $|I|=m$ ($1\le m\le M$), and denote elements of $I$ by $i(j)$, $1\le j \le m$. Then
\bea
\label{eq:mid}
I\left(X_{I};Z_{I}|Z_{I^c},S\right)
&=&
\sum_{j=1}^m
I\left(X_{i(j)};Z_{i(j)}|Z_{\{1,...,M\}\setminus \{i(j:m)\}},S\right).
\eea
\end{corollary}

Further,  suppose $\Ihat =\{1,...,m\}$ for some $2\le m\le M$. For the choice $I= \{1,...,m-1\}$ and $I' = \{m\}$, (\ref{eq:rule1'}) becomes
\be
\label{eq:cc1}
I\left(X_{\{1,...,m\}};Z_{\{1,...,m\}}|S\right)=
I\left(X_{\{1,...,m-1\}};Z_{\{1,...,m-1\}}|S\right)
+I\left(Y_m;Z_m|Z_{\{1,...,m-1\}},S\right),
\ee
which provides a useful chain rule. Applying this repeatedly, we obtain the following.

\begin{corollary}
\label{cor:chain2}
For any $1\le m \le M$,
\be
\label{eq:cc2}
I\left(X_{\{1,...,m\}};Z_{\{1,...,m\}}|S\right)=
\sum_{i=1}^{m} I\left(X_i;Z_i|Z_{\{1,...,i-1\}},S\right).
\ee
\end{corollary}

In fact, Corollary \ref{cor:chain2} can be further generalized as follows.
For any $1\le m < M$, set $\Ihat = \{1,...,M\}$, $I=\{1,...,m\}$ and $I'=\{m+1,...,M\}$ in Lemma \ref{cor:chain1} to obtain
\be
\label{eq:cc4}
I\left(X_{\{1,...,M\}};Z_{\{1,...,M\}}|S\right)
= I\left(X_{\{1,...,m\}};Z_{\{1,...,m\}}|S\right)
+
I\left(X_{\{m+1,...,M\}};Z_{\{m+1,...,M\}}|
Z_{\{1,...,m\}},S\right).
\ee
Expanding $I\left(X_{\{1,...,M\}};Z_{\{1,...,M\}}|S\right)$ and
$I\left(X_{\{1,...,m\}};Z_{\{1,...,m\}}|S\right)$ using Corollary \ref{cor:chain2},
from (\ref{eq:cc4}) we obtain the following.

\begin{corollary}
\label{cor:chain3}
For any $1\le m < M$,
\be
\label{eq:cc3}
I\left(X_{\{m+1,...,M\}};Z_{\{m+1,...,M\}}|
Z_{\{1,...,m\}},S\right)=
\sum_{i=m+1}^{M} I\left(X_i;Z_i|Z_{\{1,...,i-1\}},S\right).
\ee
\end{corollary}

\begin{lemma}
\label{le:last}
For any set $I\subseteq \{1,...,M\}\setminus \emptyset$,
\be
\label{eq:last}
I\left(X_{I};Z_{I}|Z_{I^c},S\right) \le
\sum_{i\in I} I\left(X_i;Z_i|Z_{\{1,...,i-1\}},S\right).
\ee
\end{lemma}

{\bf {\em Proof}:} Denote $m=|I|$, and let the elements $i(1),i(2),...,i(m)$ of $I$ be arranged in ascending order. Consequently, note
\be
\label{eq:last1}
\{1,...,i(j)-1\} \subseteq  \{1,...,M\}\setminus \{i(j:m)\}, \quad 1\le j \le m.
\ee
Therefore, we have
\be
\label{eq:lopP}
\{1,...,M\}\setminus \{i(j:m)\} = \{1,...,i(j)-1\} \cup \Itil(j),
\ee
where
$$\Itil(j) = (\{1,...,M\}\setminus \{i(j:m)\})\setminus \{1,...,i(j)-1\}, \quad 1\le j \le m.$$
In view of (\ref{eq:lopP}), we can write
\bea
\nonumber
&&
\!\!\!\!\!\!\!
\!\!\!\!\!\!\!
\!\!\!\!\!\!\!
I\left(X_{i(j)};Z_{i(j)}|Z_{\{1,...,M\}\setminus \{i(j:m)\}},S\right)
~~=~~ I\left(X_{i(j)};Z_{i(j)}|Z_{\{1,...,i(j)-1\}}, Z_{\Itil(j)},S\right)\\
\nonumber
&&=~~ H\left(Z_{i(j)}|Z_{\{1,...,i(j)-1\}}, Z_{\Itil(j)},S\right)
-
H\left(Z_{i(j)}|X_{i(j)},Z_{\{1,...,i(j)-1\}}, Z_{\Itil(j)},S\right)\\
\label{eq:last2}
&&=~~
H\left(Z_{i(j)}|Z_{\{1,...,i(j)-1\}}, Z_{\Itil(j)},S\right)
-
H\left(Z_{i(j)}|X_{i(j)},Z_{\{1,...,i(j)-1\}}, S\right)\\
\label{eq:last3}
&&\le~~
H\left(Z_{i(j)}|Z_{\{1,...,i(j)-1\}}, S\right)
-
H\left(Z_{i(j)}|X_{i(j)},Z_{\{1,...,i(j)-1\}}, S\right)\\
\label{eq:last4}
&&=~~ I\left(X_{i(j)};Z_{i(j)}|Z_{\{1,...,i(j)-1\}}, S\right).
\eea
Here (\ref{eq:last2}) follows by noting
$$
H\left(Z_{i(j)}|X_{i(j)},Z_{\{1,...,i(j)-1\}}, Z_{\Itil(j)},S\right)
= H\left(Z_{i(j)}|X_{i(j)}\right)
= H\left(Z_{i(j)}|X_{i(j)},Z_{\{1,...,i(j)-1\}},S\right)
$$
due to the fact that $Z_{i(j)} \rightarrow X_{i(j)}
\rightarrow (Z_{\{1,...,i(j)-1\}}, Z_{\Itil(j)},S)$ forms a Markov chain. Further, (\ref{eq:last3}) follows because conditioning reduces entropy. Now summing (\ref{eq:last4}) over $1\le j\le  m$, we obtain
\be
\label{eq:last5}
\sum_{j=1}^m I\left(X_{i(j)};Z_{i(j)}|Z_{\{1,...,M\}\setminus \{i(j:m)\}},S\right)
\le \sum_{j=1}^m I\left(X_{i(j)};Z_{i(j)}|Z_{\{1,...,i(j)-1\}}, S\right).
\ee
By Corollary \ref{cor:mid}, the left-hand side of (\ref{eq:last5}) equals $I\left(X_{I};Z_{I}|Z_{I^c},S\right)$. Also, note that the right-hand side of (\ref{eq:last5}) is same as the right-hand side of (\ref{eq:last}).
Hence (\ref{eq:last5}) is the desired result. \hfill $\Box$

{\bf {\em Proof of Lemma \ref{le:unique}}:} It is enough to show the following: if we have $I\setminus I'\ne \emptyset$ as well as $I'\setminus I\ne\emptyset$, then there exists no rate $M$-vector $R_{\{1,...,M\}}\in \Bcal^*$ such that (\ref{eq:I}) and (\ref{eq:I'}) hold simultaneously. To prove this, first we assume that (\ref{eq:I}) and (\ref{eq:I'}) hold for some $R_{\{1,...,M\}}\in \Bcal^*$ and some $(I,I')$ with the aforementioned property, and then detect a contradiction.

First consider the case where $I$ and $I'$ are disjoint.
Using (\ref{eq:rule'}) in (\ref{eq:rule1}), we obtain
\bea
\nonumber
I\left(X_{I\cup I'};Z_{I\cup I'}|Z_{{(I\cup I')}^c},S\right)
&=& I\left(X_{I};Z_{I}|Z_{{I}^c},S\right)+
I\left(X_{I'};Z_{I'}|Z_{{I'}^c},S\right)\\
\label{eq:suru1}
&&
\qquad +I\left(Z_I;Z_{I'}|Z_{{(I\cup I')}^c},S\right).
\eea
Now, adding (\ref{eq:I}) and (\ref{eq:I'}) and comparing with (\ref{eq:suru1}), we have
\bea
\nonumber
\sum_{i\in I\cup I'} R_i &=&
I\left(X_{I\cup I'};Z_{I\cup I'}|Z_{{(I\cup I')}^c},S\right)
- I\left(Z_I;Z_{I'}|Z_{{(I\cup I')}^c},S\right)\\
\label{eq:suru2}
&<& I\left(X_{I\cup I'};Z_{I\cup I'}|Z_{{(I\cup I')}^c},S\right),
\eea
because $(Z_I,(Z_{{(I\cup I')}^c},S), Z_{I'})$ does not form a Markov chain.
Note that (\ref{eq:suru2}) contradicts condition (\ref{eq:Rdef}) with $I\cup I'$ playing the role of $I$.

Next consider the case where $I\cap I' = \Itil \ne \{\}$. Writing $I=(I\setminus \Itil)\cup \Itil$, from (\ref{eq:I}), we have
\bea
\nonumber
\sum_{i\in I\setminus \Itil} R_i + \sum_{i\in \Itil} R_{i} &=& I\left(X_{I};Z_I|Z_{I^c},S\right)
~~=~~I\left(X_{(I\setminus \Itil)\cup \Itil};Z_{(I\setminus \Itil)\cup \Itil}|Z_{({(I\setminus \Itil)\cup \Itil})^c},S\right)
\\
\label{eq:suru3}
&=&
I\left(X_{I\setminus \Itil};Z_{I\setminus \Itil}|Z_{{(I\setminus \Itil)}^c},S\right) + I\left(X_{\Itil};Z_{\Itil}|Z_{\Itil^c},S\right)
+I\left(Z_{I\setminus \Itil};Z_{\Itil}|Z_{I^c},S\right)\quad \qquad
\eea
in the same manner as (\ref{eq:suru1}) with $(I\setminus\Itil,\Itil)$ playing the role of $(I,I')$. Further, from (\ref{eq:Rdef}), note that
\be
\label{eq:suru4}
\sum_{i\in \Itil} R_{i} \ge I\left(X_{\Itil};Z_{\Itil}|Z_{\Itil^c},S\right).
\ee
Using (\ref{eq:suru4}) in (\ref{eq:suru3}), we have
\be
\label{eq:suru5}
\sum_{i\in I\setminus \Itil} R_i = \sum_{i\in I\setminus I'} R_i
\le
I\left(X_{I\setminus \Itil};Z_{I\setminus \Itil}|Z_{{(I\setminus \Itil)}^c},S\right) + I\left(Z_{I\setminus \Itil};Z_{\Itil}|Z_{I^c},S\right).
\ee
Adding (\ref{eq:suru5}) and (\ref{eq:I'}), we obtain
\bea
\nonumber
\sum_{i\in I\setminus I'} R_i + \sum_{i\in I'} R_i &=& \sum_{i\in I\cup I'} R_i \\
\nonumber
&\le&
I\left(X_{I\setminus \Itil};Z_{I\setminus \Itil}|Z_{{(I\setminus \Itil)}^c},S\right) + I\left(Z_{I\setminus \Itil};Z_{\Itil}|Z_{I^c},S\right)
+ I\left(X_{I'};Z_{I'}|Z_{{I'}^c},S\right)
\\
\nonumber
&\le&
I\left(X_{I\cup I'};Z_{I\cup I'}|Z_{{(I\cup I')}^c},S\right)\\
\label{eq:suru6}
&&\qquad
- I\left(Z_{I\setminus \Itil};Z_{I'}|Z_{{(I\cup I')}^c},S\right)
+
I\left(Z_{I\setminus \Itil};Z_{\Itil}|Z_{I^c},S\right),
\eea
where the last step follows by comparing with (\ref{eq:suru1}), and letting $(I\setminus\Itil,I')$ play the role of $(I,I')$.
Further, expand
\bea
\label{eq:suru7}
I\left(Z_{I\setminus \Itil};Z_{I'}|Z_{{(I\cup I')}^c},S\right)
&=& H\left(Z_{I\setminus \Itil}|Z_{{(I\cup I')}^c},S\right)
- H\left(Z_{I\setminus \Itil}|Z_{I'\cup{(I\cup I')}^c},S\right)\\
\label{eq:suru8}
I\left(Z_{I\setminus \Itil};Z_{\Itil}|Z_{I^c},S\right)
&=&
H\left(Z_{I\setminus \Itil}|Z_{I^c},S\right)
- H\left(Z_{I\setminus \Itil}|Z_{\Itil\cup I^c},S\right).
\eea
Now, note $I'\cup{(I\cup I')}^c = \Itil\cup I^c$, and subtract (\ref{eq:suru8}) from (\ref{eq:suru7}) to obtain
\bea
\nonumber
&&
%\!\!\!\!\!\!\!\!\!\!
\!\!\!\!\!
\!\!\!\!\!
I\left(Z_{I\setminus \Itil};Z_{I'}|Z_{{(I\cup I')}^c},S\right)
- I\left(Z_{I\setminus \Itil};Z_{\Itil}|Z_{I^c},S\right)\\
\nonumber
&&=~~ H\left(Z_{I\setminus \Itil}|Z_{{(I\cup I')}^c},S\right)
- H\left(Z_{I\setminus \Itil}|Z_{I^c},S\right)\\
\label{eq:suru9}
&&=~~ I\left(Z_{I\setminus \Itil};Z_{I'\setminus \Itil}|Z_{{(I\cup I')}^c},S\right)\\
\label{eq:suru10}
&&>~~ 0.
\eea
Here (\ref{eq:suru9}) follows by noting $I^c= {(I\cup I')}^c \cup (I'\setminus \Itil)$ with ${(I\cup I')}^c$ and $I'\setminus \Itil$ disjoint. Further, (\ref{eq:suru10}) follows due to the fact that $(Z_{I\setminus \Itil}, (Z_{{(I\cup I')}^c},S), Z_{I'\setminus \Itil})$ do not form a Markov chain. Using (\ref{eq:suru10}) in (\ref{eq:suru6}), we again obtain (\ref{eq:suru2}), which as earlier contradicts (\ref{eq:Rdef}). Hence the result. \hfill $\Box$

\end{document}